\documentclass[iop,apj]{emulateapj}

\usepackage{multirow}
\usepackage{hyperref}

\def\xmm{{XMM-{\it Newton}}}
\def\chandra{{\it Chandra}}

\newcommand{\cgs}{ ${\rm erg~cm}^{-2}~{\rm s}^{-1}$} 

\def\gtrsim{\mathrel{\hbox{\rlap{\hbox{\lower4pt\hbox{$\sim$}}}\hbox{$>$}}}}

\usepackage{txfonts}
\usepackage{natbib}

\bibpunct{(}{)}{;}{a}{}{,}
\slugcomment{Accepted to the Astrophysical Journal on October 3, 2013}

\shorttitle{Elusive inverted P-Cygni in CID-42}
\shortauthors{Lanzuisi et al.}

\begin{document}


\title{The XMM-Newton spectrum of a candidate recoiling supermassive black hole: an elusive inverted P-Cygni profile}

\author{G. Lanzuisi\altaffilmark{1,2}, F. Civano\altaffilmark{1,3}, S. Marchesi\altaffilmark{1,4}, 
A. Comastri\altaffilmark{5}, E. Costantini\altaffilmark{6}, M. Elvis\altaffilmark{3}, V. Mainieri\altaffilmark{7}, R. Hickox\altaffilmark{1}, K. Jahnke\altaffilmark{8}, S. Komossa\altaffilmark{9}, E. Piconcelli\altaffilmark{10}, C. Vignali\altaffilmark{4}, M. Brusa\altaffilmark{4}, N. Cappelluti\altaffilmark{5} and A. Fruscione\altaffilmark{3}} 

\altaffiltext{1}{Department of Physics and Astronomy, Dartmouth College, Wilder Laboratory, 03855 Hanover, NH, USA}
\altaffiltext{2}{Institute of Astronomy \& Astrophysics, National Observatory of Athens, Palaia Penteli, 15236, Athens, Greece}
\altaffiltext{3}{Smithsonian Astrophysical Observatory, 60 Garden st., 02138 Cambridge, MA, USA}
\altaffiltext{4}{Dipartimento di Astronomia, Universit\'a degli Studi di Bologna, via Ranzani 1, 40127, Bologna, Italy}
\altaffiltext{5}{INAF- Osservatorio Astronomico di Bologna, via Ranzani 1, 40127, Bologna, Italy}
\altaffiltext{6}{SRON, Netherlands Institute for Space Research, Sorbonnelaan, 2, 3584 CA, Utrecht, The Netherlands}
\altaffiltext{7}{European Southern Observatory, Karl-Schwarschild-Strasse 2, 85748, Garching bei Munchen, Germany}
\altaffiltext{8}{Max Planck Institute for Astronomy, Konigstuhl 17, D-69117 Heidelberg, Germany}
\altaffiltext{9}{Max-Planck-Institut fuer Radioastronomie, Auf dem Huegel 69, 53121 Bonn, Germany}
\altaffiltext{10}{INAF-Osservatorio Astronomico di Roma, via Frascati 33, 00040 Monteporzio Catone, Italy}


\begin{abstract}
We present a detailed spectral analysis of new \xmm\ data of the source CXOC J100043.1+020637, also known as CID-42, detected in the COSMOS survey 
at z = 0.359. Previous works suggested that CID-42 is a candidate recoiling supermassive black holes showing also an {\it inverted} P-Cygni profile in the X-
ray spectra at $\sim$6 keV (rest) with an iron emission line plus a redshifted absorption line (detected at 3 $\sigma$ in previous \xmm\ and  \chandra\ 
observations). Detailed analysis of the absorption line suggested the presence of ionized material inflowing into the black hole at high velocity. In the new 
long \xmm\ observation, while the overall spectral shape remains constant, the continuum 2-10 keV flux decreased of $\sim$20\% with respect to previous 
observation and the absorption line is undetected. The upper limit on the intensity of the absorption line is EW$<$162 keV. Extensive Monte Carlo 
simulations show that the non detection of the line is solely due to variation in the properties of the inflowing material, in agreement with the transient nature 
of these features, and that the intensity of the line is lower than the previously measured with a probability of 98.8\%. In the scenario of CID-42 as recoiling 
SMBH, the absorption line can be interpreted as due to inflow of gas with variable density and located in the proximity of the SMBH and recoiling with it.  New 
monitoring observations will be requested to further characterize this line.
\end{abstract}

\section{Introduction}
During galaxy major mergers, the central supermassive black holes (SMBHs) that reside within the merging galaxies will 
form a bound binary that can itself merge (e.g. Volonteri et al. 2003, Hopkins et al. 2008, Colpi \&\ Dotti 2009). 
At the time of the SMBH binary coalescence, strong gravitational wave (GW) radiation is emitted an-isotropically, depending on the spin and mass-ratio of 
the two SMBHs. To conserve linear momentum, the newly formed single SMBH recoils (Peres 1962, Bekenstein 1973).
Recoiling SMBHs are the direct products of processes in the strong field regime of gravity and are one of the key observable signatures of a SMBH binary 
merger.  As the SMBH recoils from the center of the galaxy, the closest regions (disk and broad line regions) are carried with it and the more distant region 
are left behind depending on the recoil velocity (Merritt et al. 2006, Loeb 2007). Because GW recoil displaces (or ejects) SMBHs from the centers of 
galaxies, these events have the potential to influence the observed 
co-evolution of SMBHs with their host galaxies, as demonstrated by numerical simulations (Blecha et al. 2011, Sijacki et al. 2011, Guedes et al. 2011). 
Observational searches for recoiling SMBHs are just at the beginning (Bonning et al. 2007, Eracleous et al. 2012, see Komossa 2012 for a review) and only 
few serendipitous discoveries of candidates have been reported in the literature (Komossa et al. 2008, Shields et al. 2009, Robinson et al. 2010, Jonker et 
al. 2010, Batcheldor et al. 2010, Steinhardt et al. 2012) . 

The \chandra-COSMOS source \allowbreak{CXOC J100043.1+020637} ($z$=0.359, Elvis et al. 2009, Civano et al. 2012), also known as CID-42, is a 
candidate for being a GW recoiling SMBH with both imaging (in optical and X-ray) and spectroscopic signatures (Civano et al. 2010, 2012; hereafter C10 and 
C12). The current data are consistent with a recoiling SMBH ejected $\sim$1-6 Myr ago, as shown by detailed modeling presented in Blecha et al. (2013). 

\begin{figure*}
\begin{center}
\includegraphics[width=0.5\textwidth,angle=-90]{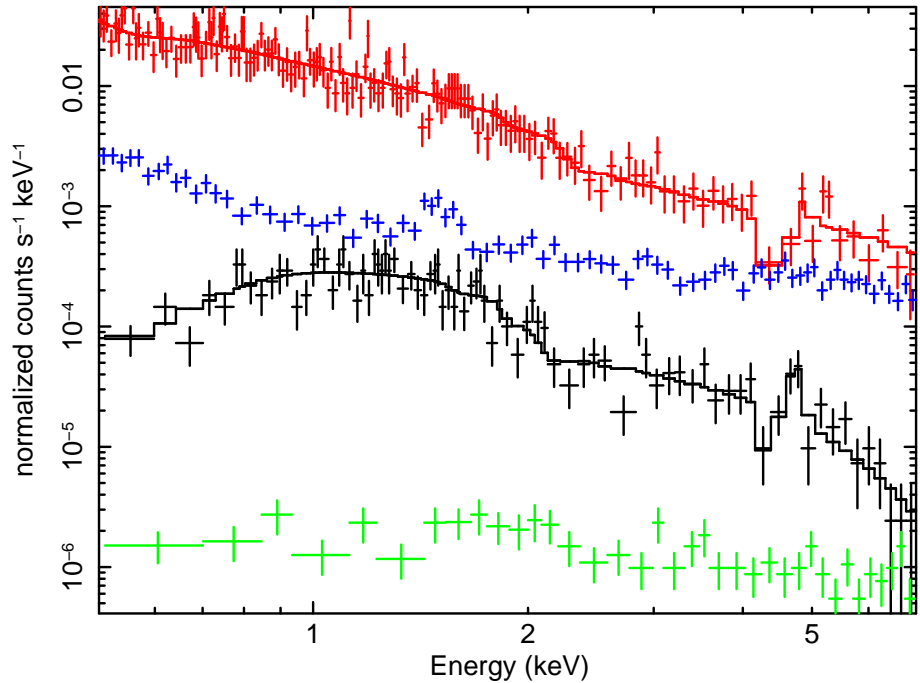}

\end{center}
\caption{EPIC pn (red) and ACIS (black) total spectra of CID-42 adapted from Figure 7 of C10 and rebinned for plotting purposes. The background spectra (pn in blue and ACIS in green) are reported and rescaled to the source area. The y scale is arbitrary but the ratio between source and background has been preserved. The best fit model (absorbed power-law, thermal component and emission and absorption features) is plotted.  }\label{oldspec}
\end{figure*}

\begin{figure*}
\begin{center}
\includegraphics[width=0.5\textwidth]{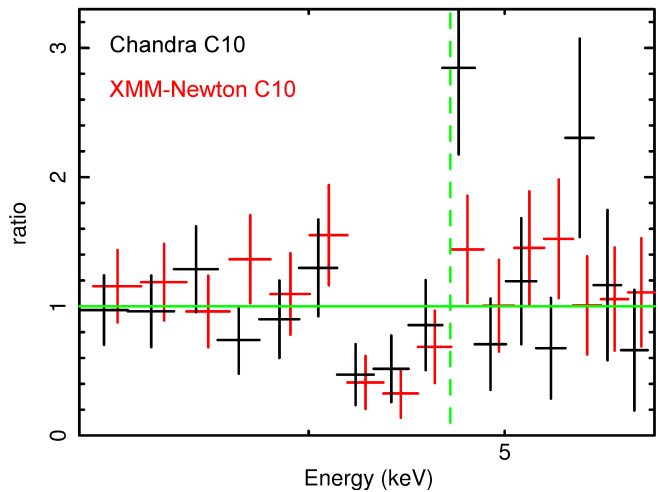}

\end{center}
\caption{Data to model ratio around the energy where the inverted {\it P-Cygni} profile is detected in EPIC pn (red) and ACIS (black) spectra of CID-42 adapted from C10. The model used here does not includes the emission and absorption features, only the continuum model is used. The vertical line corresponds to the 6.4 keV rest frame energy. The spectra have been rebinned in energy with $\Delta$E$\sim$200 eV in order to highlight the absorption features. }\label{oldspecratio}
\end{figure*}

In both the \chandra\ and \xmm\ spectra of CID-42, a remarkable inverted {\it P-Cygni} profile, i.e., an absorption feature redshifted with respect to the 
emission component, was detected by C10 at  $\sim$ 4.5 keV in the observed frame and $\sim$6 keV in the rest frame (Fig. \ref{oldspec} and \ref
{oldspecratio} adapted from Fig. 7 of C10).
The emission feature was consistent with being a neutral iron line at the system redshift, with a constant flux but more prominent in the \chandra\ spectrum 
(EW = 570 $\pm$ 260 eV) than in the \xmm\ one (EW=142$^{+143}_{-86}$eV). 
The absorption feature (at $\sim$6 keV in the rest frame), detected in both \xmm-EPIC pn and \chandra-ACIS spectra, showed a line energy centroid 
changing in time between 5.8 and 6.2 keV ($\Delta$E$_{rest}\sim$ 500 eV; see Figure 8 of C10), with intensity of 350$\pm$120 eV.
A re-analysis of C10 data, including extensive Monte Carlo simulations (see Section \ref{oldsim}), shows that the significance of the redshifted absorption line 
in the \xmm\ data is 3$\sigma$, strengthened by the detection of the line also in \chandra-ACIS at 2.2$\sigma$ (Figure \ref{oldspecratio}).

Most known X-ray absorbers in Active Galactic Nuclei (AGN) are observed as blueshifted lines and so are signatures of fast outflowing winds (see Tombesi 
et al. 2013 and references therein) predicted in both phenomenological and semi-analytical quasar models (e.g., Elvis 2000, King \& Pounds 2003, King 
2010). Redshifted absorbers, instead, would seem to require high-velocity inflows, which must therefore be located in the proximity of the SMBH. 


Few cases of objects with redshifted absorption lines are reported in the literature (NGC 3516, Nandra et al. 1999; E1821+643, Yaqoob \& Serlemitsos 2005; Mrk 509, Dadina et al. 2005; PG 1211+143, Reeves et al. 2005; Q0056-363, Matt et al. 2005; Ark 120, Nandra et al. 2007; Mrk 335, Longinotti et al. 2007). In most cases, the redshifted absorption line has not been observed in additional observations (Vaughan \&\ Uttley 2008 for an extended discussion on the statistical significance of these lines). 
As for example, the redshifted absorption line in Mrk 335 was reported by Longinotti et al. (2007). Later paper with new data for this source do not search for the redshifted absorption line previously detected (O' Neill et al. 2007, Grupe et al. 2012). The redshifted line seen in Mrk 509 with Beppo-SAX (Dadina et al. 2005) was not detected again by Cappi et al. (2009) in 5 \xmm\ observations taken between 2000 and 2006. Ponti et al. (2013) observed again Mrk 509 in 2009 as part of a long \xmm\ campaign (10 $\times$ 60 ks observations) but the line was reported as undetected again. 
In PG 1211+143, a redshifted absorption line was first observed (Reeves et al. 2005) however in newer data a classic {\it P-Cygni} profile only is reported (Pounds et al. 2009). Q0056-363 was observed in 2000 and in 2003: both the observations are discussed in Matt et al. (2005). The redshifted absorption line was detected only in 2003 data. Moreover, this observation has been split in five time-intervals in order to study short time-scale variability: the absorption line was found in only one of these intervals. Tombesi et al. (2010) also reported the detection (at a 90\% confidence level using the F-test) of redshifted absorption lines in seven more sources (including Mrk 335). 
All these evidences suggest that these features are highly variable, and that their occurrence, or duty cycle, is very low. 

The most convincing case is E1821+643, a candidate recoiling SMBH (Robinson et al., 2010), like CID-42, with also an iron redshifted absorption line in its \chandra\ X-ray spectrum, observed by Yaqoob \& Serlemitsos (2005). E1821+643 was observed again with \xmm\ by Jim\'enez-Bail\'on et al. (2007), who reported a new measure for the peak energy of the absorption line. The absorption line has been explained as gravitationally redshifted iron absorption and its line energy variability as changes in the ionization state of the iron clouds.

C10 suggested a possible interpretation for the redshifted absorption line seen in CID-42 as gas (either neutral or ionized iron) infalling into the recoiling SMBH at relativistic velocities (0.02-0.14$c$), but, given the data quality and the degeneracy between velocity and ionization state, a firm conclusion on the properties of the infalling material could not be reached.

In order to perform an accurate characterization of the X-ray absorber (density, velocity, covering factor and ionization state), we requested and obtained a 123 ks long un-interrupted \xmm\ observations. 
In this paper, we present the analysis of the line profile as well as the broad band spectral shape and the obtained results.

Throughout the paper 90\% errors (1.6 $\sigma$) are quoted unless otherwise stated. 


\begin{figure*}
\begin{center}
\includegraphics[width=0.5\textwidth]{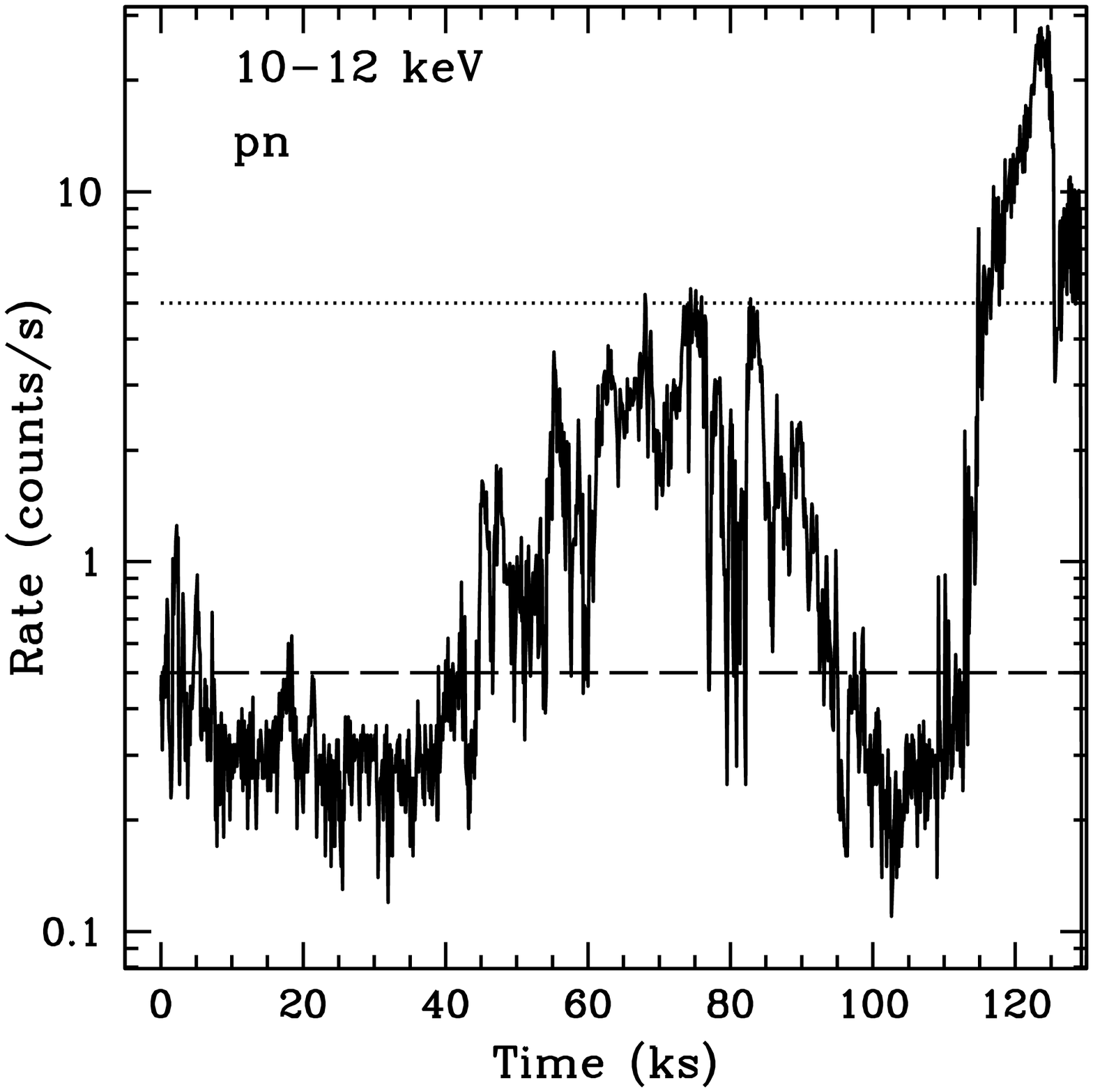}
\end{center}
\caption{The 10--12 keV light curve extracted from the pn data. The dotted and dashed lines represents the cuts applied to the data in case A and B respectively.}\label{lc}
\end{figure*}

\section{Data reduction}

CID-42 was observed by \xmm\ with a 123 ks single observation taken on June 3, 2011 (during revolution 2103, observation ID 0672780101, PI: F. Civano). Standard \xmm\ SAS tasks \textit{epproc} and \textit{emproc} (SAS 12.0.1) were used to produce calibrated pn and MOS event file. These event files were processed using the latest calibration database and cleaned up from hot pixels and cosmic rays contamination. X-ray events corresponding to patterns 0-4 (0-12) for the pn (MOS) cameras were selected. 

A global light-curve at energies greater than 10 keV (where the contribution from the emission of X-ray sources is negligible) was produced in order to identify any background flare (Figure \ref{lc}). 
The last $\sim$15 ks of the observation are affected by high background ($\sim$25 ct/s), typical of the edge of each \xmm\ observing orbit.
Moreover, the central part of the observation has a long ($\sim$ 50 ks) interval during which the background intensity is 10 times stronger than the remaining time in this observation.

For the purposes of this work (i.e., study the absorption feature), two different approaches have been used in filtering the data to remove the flare: (A) to maximize the number of counts, (B) to maximize the signal-to-noise ratio. 
In the less conservative case (A), a standard 3$\sigma$ clipping method has been used to discard intervals with high background rate (dotted line in Fig. \ref{lc}).The part of the observation where the background contribution is lower than 5 cts s$^{-1}$ (0.5 cts s$^{-1}$) in the pn (MOS) has been included. In the more conservative case (B), only the time intervals with background intensity $<$0.5 ($<$0.15) cts s$^{-1}$ in the pn (MOS) have been considered for the analysis (dashed line in Fig. \ref{lc}), obtained by applying twice a 3$\sigma$ clipping cut. 
This level is consistent with the one used in Cappelluti et al. (2007) to clean the \xmm\ data used in C10 analysis. After the cleaning operation, a net exposure time of $\sim$100 ($\sim$110) ks and $\sim$53 ($\sim$80) ks is available for the pn (MOS) camera in case A and B, respectively (see Table  \ref{tab:counts}). 

Source and background spectra were extracted for both case A and B from the data of each camera. 
The source spectra were extracted from circular regions of radius 25$^{\prime\prime}$ (MOS1 and MOS2) and 30$^{\prime\prime}$ (pn), maximizing the number of source counts and minimizing the background contribution.
The background spectra were extracted from multiple regions close to the target, avoiding CCD gaps and other X-ray detected sources listed in the COSMOS X-ray catalogs (Cappelluti et al. 2009, Brusa et al. 2010, Elvis et al. 2009, Civano et al. 2012). The background extraction areas are typically about 20 times the source extraction region, in order to average out all the position-dependent background features and to obtain enough background counts and perform accurate spectral analysis. The background spectra are shown in Figure \ref{newspec} as green and blue points. 

Table \ref{tab:counts} lists net counts in each camera in 3 bands (0.5-7, 0.5-2 and 2-7 keV), together with the same numbers for the pn spectrum reported in C10. In the same table we also report the fraction of source counts in the extraction region in the 0.5-7 keV band (column 5) and also in the 2-7 keV band (column 6). The source signal-to-noise ratio in the 2-7 keV band, where the absorption and emission lines were previously observed, is lower in case A spectra than in B, where the high background flare was removed. This effect is very important in the pn with a difference of $\sim$30\% between A and B in the 2-7 keV band. The net number of counts however is higher in case A than in case B. In the next section, a discussion on the spectra choice is presented.



\begin{table*}[t]
\centering
\caption{Exposure times and net counts in the three cameras after the cleaning operations. For a comparison, we also report the same values for the observation used in C10.}
\begin{tabular}{lccccccc}
\hline 
\hline
&Instrument & Exposure time & Counts & Counts & Counts & Source \% &   Source \%\\
                 &  & ks   &  0.5-7 keV & 0.5-2 keV & 2-7 keV & 0.5-7 keV & 2-7 keV \\
\hline
\multirow{2}{*}{A} &MOS1+MOS2 & 225.3 & 1950 & 1539 & 404 &84.4 & 66.8 \\ 
&pn & 100.8  & 3013 & 2502  &  511 & 77.9 & 49.0\\

\hline
\multirow{2}{*}{B} &MOS1+MOS2 & 160.3 &1420 & 1106 & 314 &86.8 & 75.4 \\ 
&pn & 52.77 & 1622 & 1370 & 251 & 87.2 & 68.1\\
\hline
C10 & pn  & 71.2 & 1500 & 1237 & 280 & 87.1  & 76.4 \\
\hline
\hline
\end{tabular}
\label{tab:counts}
\end{table*}

\section{Spectral analysis}

To perform the spectral analysis, we followed the approach of C10, where the spectral fit is performed on the total spectrum, modeling the source and the background simultaneously (see Fiore et al. 2012 and Lanzuisi et al. 2013 for details on this approach), after first determining an accurate model for the background.

The \xmm\ EPIC pn and MOS background spectra ($\sim$8000 and $\sim$4000 counts each in case A and B) were fitted over the 0.3-7 keV range using XSPEC version 12.8 (Arnaud 1999) using the following components: two power-law components, a thermal component for the soft part of the spectrum, 3 Gaussian emission lines to reproduce the features of the pn and MOS backgrounds\footnote{See the list of emission lines for both the pn and MOS camera at this web site \url{http://www.star.le.ac.uk/\~amr30/BG/BGTable.html}.}. This background best-fit model was then rescaled to the source area using the ratio of the extraction region areas (BACKSCAL keyword).

\begin{figure*}
\begin{center}
\includegraphics[width=0.5\textwidth,angle=-90]{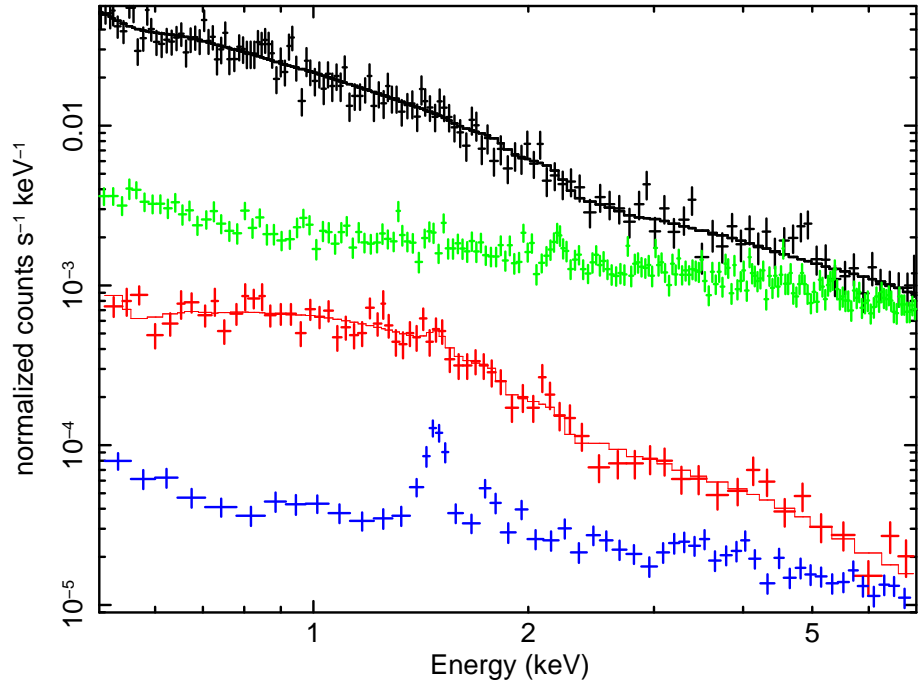}
\end{center}
\caption{EPIC pn (black) and MOS1$+$MOS2 (red) total spectra (case A), rebinned for plotting purposes. The background spectra (pn in green and MOS in blue) are reported and rescaled to the source area. The y-axis normalization is arbitrary but the source to background ratio is preserved. The model is the best fit reported in Table 2 (no emission line is included).  }\label{newspec}
\end{figure*}

The source spectra were fitted in the 0.5-7 keV band with an absorbed power-law plus a thermal component ({\it mekal}), the same model used in C10. A Galactic column density along the line of sight of N$_{H,Gal}$=2.6$\times$10$^{20}$ cm$^{-2}$ (Kalberla et al. 2005) has been included as well. The modified Cash statistic  implemented in XSPEC ({\it cstat}; Cash 1979) has been employed for the fitting, and a minimum binning of 1 count per bin has been applied to the spectra. 
The pn and MOS1+MOS2 spectra were first fitted separately to verify the consistency of the fit between instruments, then jointly to more tightly constrain the errors on the spectral parameters. In Figure \ref{newspec}, the case A spectra are shown (top: pn with background spectrum; bottom: MOS with background spectrum). The best fits of the spectra in case A and B are fully consistent within the errors, however, in case A, where the number of source counts is higher, the parameters are better constrained. 

The spectral analysis results for the joint fit are listed in Table \ref{tab:spectralanalysis} for case A together with the best fit parameters reported in C10. 
The spectral index obtained from the joint fit is $\Gamma$=2.16$\pm$0.08, slightly steeper than the slope measured in C10. The presence of a thermal component is significant at $>$5$\sigma$, as indicated by the F-test, and the temperature is 
$kT$ = 0.18 $^{+0.02}_{-0.04}$keV.  An upper limit of N$_{H}$$<$ 6$\times$10$^{20}$ cm$^{-2}$ is measured for the intrinsic absorption.

\begin{table*}[t]
\begin{center}
\caption{Spectral analysis results (continuum and emission line) and fluxes for the joint spectral fitting of pn and MOS1+MOS2 and the values from the ACIS-I and pn spectra analyzed in C10. 
Errors are at 90\% confidence level.}\label{tab:spectralanalysis}
\begin{tabular}{cccc}
\hline
\hline
Fit Parameter  & ACIS-I (C10) & pn (C10) & pn+MOS (here) \\
\hline
Continuum & & \\

$\Gamma$ &     1.88$_{-0.13}^{+0.17}$ & 1.95$_{-0.06}^{+0.07}$ &  2.16$_{-0.08}^{+0.08}$\\
$N_{H}$ (10$^{22}$ cm$^{-2}$) &     $<$0.2 &       $<$0.02 & $<$ 0.06 \\
$kT$ (keV) & $<$ 0.11 &     $<$ 0.13 &                  0.18 $^{+0.02}_{-0.04}$\\
F$_{0.5-2 keV} (10^{-14}$ \cgs)   & 1.8$\pm$0.2 & 4.8$\pm$0.3      &5.1$\pm$0.2 \\
F$_{2-10 keV} (10^{-14}$ \cgs)   &2.9$\pm$ 0.3  &   6.1$\pm$0.4   &4.9$\pm$0.3\\
F$_{0.5-10 keV} (10^{-14}$ \cgs) &5.1$\pm$ 0.4  &  10.9$\pm$0.6  &  10.1$\pm$0.5\\
\hline
Emission Line (one Gaussian fit) & & \\
Observed Energy  & 6.44$\pm$0.07 &  6.60$\pm$0.15& 6.62$_{-0.09}^{+0.12}$ \\
Line $\sigma$ (keV) & $<$0.12 & $<$0.2& $<$0.6 \\
EW (eV) & 570$\pm$260  &142$^{+143}_{-86}$eV  & 593$_{-390}^{+347}$  \\ 
\hline
Emission Line (double Gaussian fit) & & \\
Observed Energy &\dotfill& \dotfill&6.70$_{-0.13}^{+0.11}$  \\
Line $\sigma$ (keV) &\dotfill & \dotfill&$<$0.2  \\
EW (eV) &  \dotfill &\dotfill& 394$^{+278}_{-160}$ \\ 
Observed Energy & \dotfill &\dotfill&6.41$_{-0.30}^{+0.37}$  \\
Line $\sigma$ (keV) & \dotfill &\dotfill& $<$0.2 \\
EW (eV) & \dotfill &\dotfill&  $<$500 \\ 

\hline
\hline

\end{tabular}
\end{center}
\end{table*}

\begin{figure*}
\centering \includegraphics[width=0.5\textwidth]{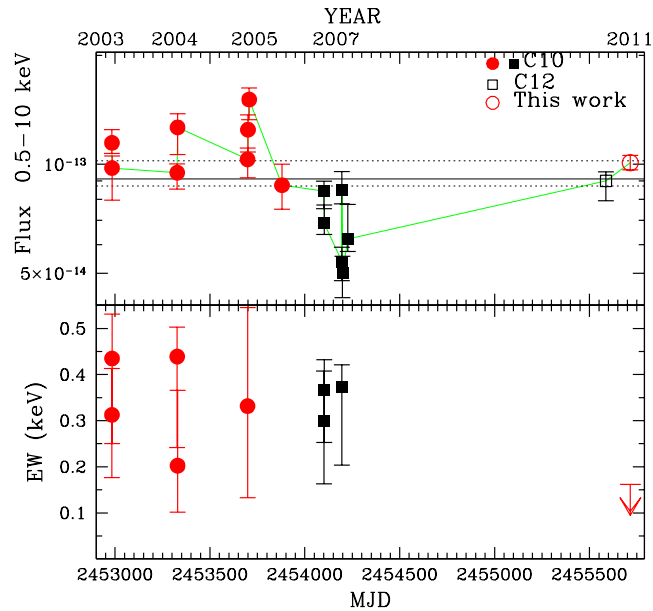}
\caption{{\it Top:} X-ray full band flux light curve (red=\xmm\, black=\chandra) adapted from C10, including the data point from the 2011 \xmm\ data and also C12 \chandra\ HRC data. The reported errors are at 90\% confidence level. The mean value and error are reported as horizontal lines. {\it Bottom:} Absorption line equivalent width measured in different epochs spectra. }\label{lightcurve}
\end{figure*}

The model fluxes, computed in the 0.5-2, 2-10 and 0.5-10 keV bands, are reported in Table \ref{tab:spectralanalysis}. The 0.5-10 keV band flux is compared in Figure \ref{lightcurve} with the fluxes obtained in the previous \xmm\ and \chandra\ observations (C10 and C12). The light curve spans a period of $\sim$9 years. The flux reported here is consistent with the flux reported in C10 for the \xmm\ data and consistent within the error with the C12 flux for the \chandra\ HRC observation taken in January 26, 2011.  
While the soft component of the spectrum remains almost constant, the hard part of the spectrum is 20\% fainter here than in C10.
The flux light curve of CID-42 has been analyzed in a variability study of the XMM-COSMOS sample by Lanzuisi et al. (submitted), where it is shown that CID-42 is among the 20\% most variable sources among the brightest ($>$1000 counts) 65 sources in their sample ($\sim$900 sources total). The light curve of the source within the 2011 \xmm\ observation studied here does not show short time scale variability.

\begin{figure*}
\begin{center}
\includegraphics[width=0.5\textwidth]{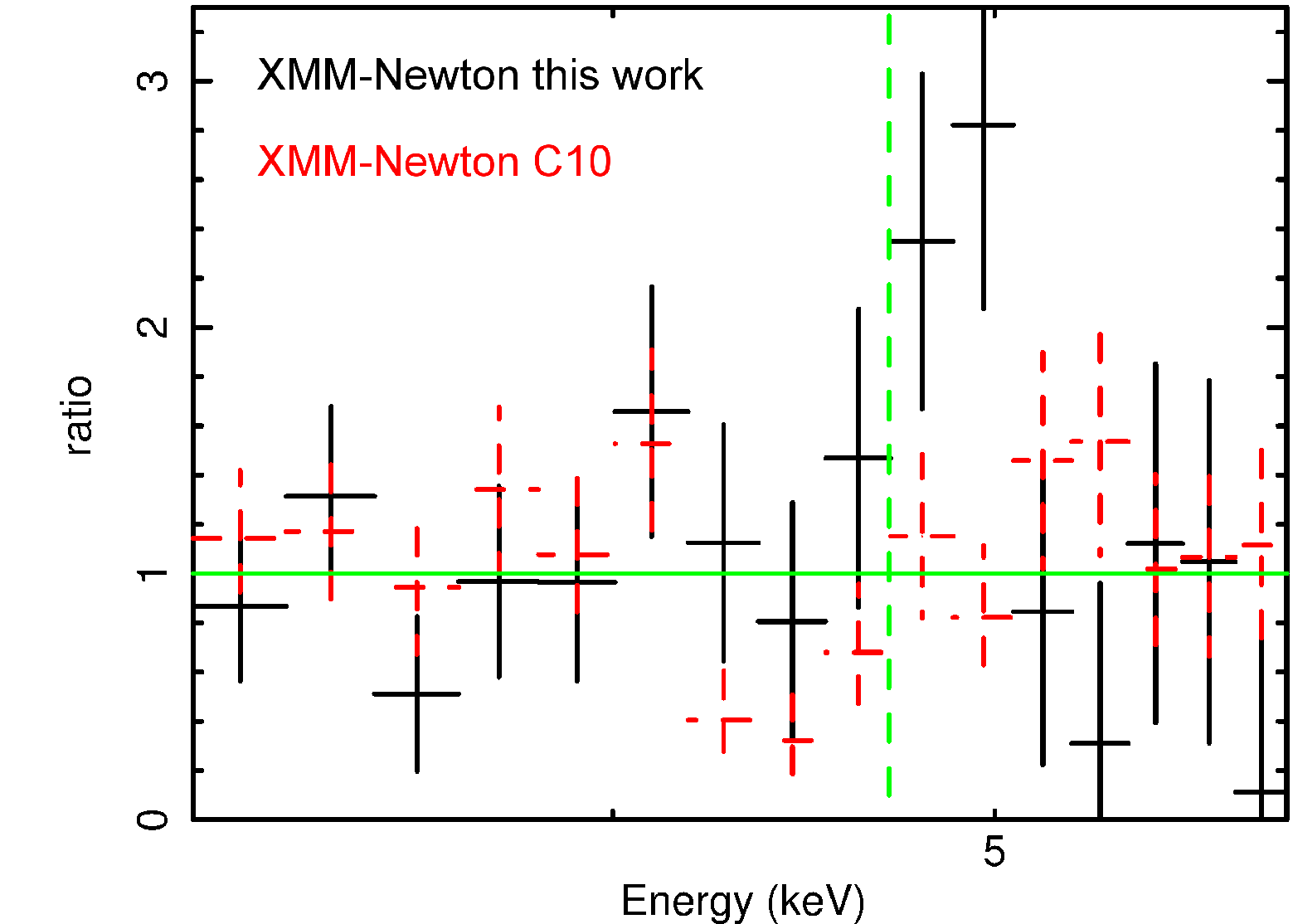}
\end{center}
\caption{A zoom on the data to model (no features included, only the continuum model is used) ratio around the energy of the inverted {\it P-Cygni} profile. C10 pn data (red dashed line) are compared to the new pn data (black solid). The vertical line corresponds to the 6.4 keV rest frame energy. The data in the zoom have been rebinned in energy with $\Delta$E=200 eV to highlight the presence of the absorption feature.}\label{oldnewzoom}
\end{figure*}

\section{P-Cygni profile analysis}

The main goal of this work is to study the properties of the absorption feature that was detected in the previous \xmm\ and \chandra\ observations (see Section \ref{oldsim}), however no clear evidence of it has been found in the new data. The iron emission line instead is still detected and prominent. 
In Figure \ref{oldnewzoom}, the ratio between data and model of the new pn data with those presented in C10 is shown (the expected energy of the 6.4 keV Fe K$\alpha$ line is plotted). 

Both the A and B spectra were used for the search of this line and both analyses do not return a detection. 
The analysis of spectrum B returns only a limit of $<$ 248 eV (90\% confidence level) on the equivalent width of the absorption line. 
The spectrum A, having 60\% more counts than spectrum B in the hard band, returns a tighter limit on the equivalent width of EW$<$162 eV.
The above upper limits have been computed at the energy where the absorption line was detected in C10 (4.5 keV observed frame).  Given that, as discussed in the introduction, the line centroid has been observed to change between 4.2 and 4.6 keV in the observed frame, it might be slightly different from the nominal value reported in C10. The upper limits fixing the line centroid in the 4-4.7 keV observed range are consistent within $\pm$20 eV (\%10) with the above values. This finding is consistent with the observed flat spectral slope measured on the broad band spectrum.
In Figure \ref{lightcurve} (bottom panel), the EW value measured is plotted as un upper limit together with the intensities measured in C10 \xmm\ and \chandra\ spectra. 

The emission line was first fitted with a narrow unresolved Gaussian line in both A and B (the numbers listed afterward are for spectrum A and are reported in \ref{tab:spectralanalysis}). The line centroid is higher than what expected for the neutral FeK$\alpha$ line at 6.4 keV: the best fit rest frame energy in the joint fit is 6.62$_{-0.09}^{+0.12}$ keV, consistent with the value reported in C10 of 6.60$_{-0.12}^{+0.15}$ keV for the \xmm\ spectrum. In the \chandra\ spectrum instead, C10 reported an energy of 6.44$\pm$0.07 keV.  
The equivalent width of the line is  593$_{-390}^{+347}$ eV (pn value), brighter than the \xmm\ value reported in C10 of 142$^{+143}_{-86}$ eV and comparable to the \chandra\ value ($EW_{Chandra}$=570$\pm$ 260 eV).

Leaving the emission line width free to vary, the profile width is $\sigma <$0.6 keV and can be fitted by using two narrow Gaussian lines, with rest frame energies of 6.70$_{-0.13}^{+0.11}$ and 6.41$_{-0.30}^{+0.37}$ keV and equivalent width of 394$^{+278}_{-160}$ eV and $<$500 eV respectively (obtained by fixing the energy at the best fit). The emission line can be interpreted as the blended emission of the 6.4 and 6.7 keV iron emission lines (neutral and He-like iron), however the two fits (one or two lines) are statistically indistinguishable.

In C10, the iron emission line is consistent with a single narrow line in both \chandra\ and \xmm\ spectra. No hint of a second emission line or of a broad profile is observed. ÊWhile the emission line in the \chandra\ spectrum is perfectly consistent with a neutral Fe K$\alpha$ line at 6.4 keV, the energy observed in the \xmm\ spectrum is higher (6.6 keV). This can be however interpreted as the effect of the broader absorption feature in \xmm\,Ê
that allows only the blue wing of the line to be observed (Fig. \ref{oldspec} and \ref{oldspecratio}).

\subsection{Statistical significance of absorption line in previous data}
\label{oldsim}

Extensive Monte-Carlo simulations (10k runs) were carried out, producing simulated spectra
with the FAKEIT routine in XSPEC, to estimate the significance of the C10 redshifted absorption line. 
An absorbed power-law with the best-fit parameters obtained from the observed data was used as input model. 
The simulated spectra were fitted first with a power-law model, then a narrow absorption Gaussian line has been added to the model. The absorption line energy has been left free to vary in the range 2--7 keV as well as the line (negative) intensity.
About 30 of the simulated spectra shows a $\Delta$Cash $>$ 12, the value observed in the C10 \xmm\ spectrum between a fit without and with line.
Therefore we estimate that the probability to detect the observed $\Delta$Cash by chance in the \xmm\ data alone is $\sim$3$\times$10$^{-3}$,  
i.e. the feature is significant at confidence level $\sim$99.7\% (3$\sigma$).
The significance is strengthened by the detection of the line also in Chandra-ACIS at 2.2$\sigma$ (Figure \ref{oldspecratio}).

\subsection{Statistical significance of the absorption feature}

To assess whether the apparent lack of the absorption line is due to the data quality rather than a real disappearance of the line itself, 
extensive Monte Carlo simulations have been carried out using the FAKEIT routine within XSPEC. 
Two different sets of spectra were simulated: one using the clean exposure time of case B and the second using the less conservative case A. The simulated spectra have the same continuum flux and background level as observed in 2011 plus an absorption iron line with the same properties of the one observed in C10 (i.e., same equivalent width, width and observed energy of 4.5 keV). 
Moreover, even though the equivalent width measured in C10 for the line was constant within error bars (Figure 8 of C10), further simulations decreasing the intensity of the line from 95\% to 5\% (in step of 5\%) of the C10 value have been simulated. 

The fake spectra were first fitted with a power-law model plus an absorption line with a centroid energy fixed at the \xmm\ C10 value and normalization free to vary to both positive and negative values. 

\begin{figure*}
\centering \includegraphics[width=0.5\textwidth]{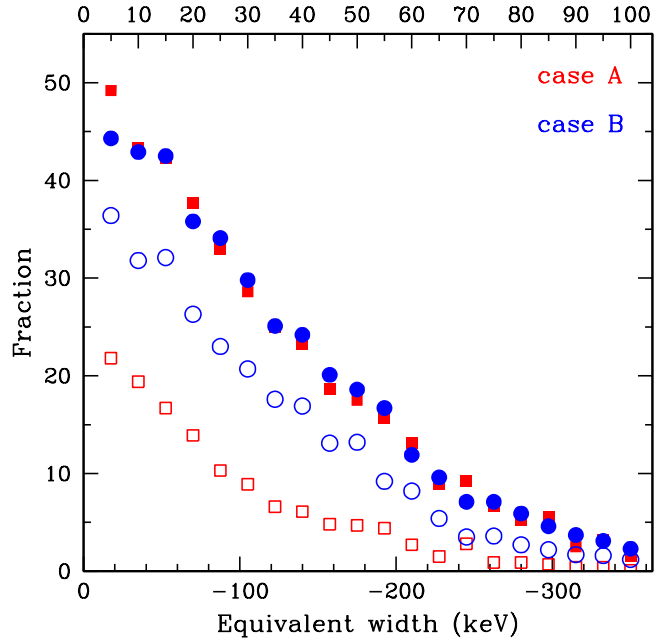}
\caption{Fraction of simulated spectra with measured line intensity consistent with zero (solid symbols) and lower than the upper limit measured in the observed spectra (open symbols) as function of the input line intensity (lower x-axis in keV and upper x-axis in fraction of the EW in C10) for case A (red squares) and B (blue circles).}\label{ew}
\end{figure*}

The intensity of the line has been measured in all the simulated spectra, as well as the 90\% confidence value. In Figure  \ref{ew}, the fraction of spectra where the measured intensity of the line is either consistent with zero or lower than the 90\% upper limit measured in the observed data have been computed. These values (solid and open symbols respectively) are reported as function of the input line intensity for case A (red) and B (blue) in both eV units (bottom x-axis) and also with respect to the value in C10 (top x-axis). 

The analysis of the simulations allows to confirm that the case A and B spectra return statistically consistent results. For line intensity lower that 70\% of C10,  the most stringent upper limit can be obtained with case A spectra (larger number of counts) regardless of higher background contribution. 

The probability that the line in the observed data has and intensity comparable with the one observed in C10 is lower than 2.3\% with a more stringent limit of 1.2\% obtained measuring the fraction of sources with 90\% confidence value lower than the one measured in these data. The probability that the line has an intensity (at 90\% confidence) comparable to 50\% of the C10 value ranges between 5 to 15\%, and it becomes higher with fainter input lines reaching 20-35\%. 

\section{Discussion}

In this paper we analyzed the most recent X-ray spectrum of CID-42 obtained during mid--2011 with the \xmm\ satellite. The goal of this long observation was to properly characterize the unprecedented redshifted, variable in energy, iron absorption line observed in previous \xmm\ and \chandra\ observations. 

Two different \xmm\ spectra have been used for the analysis: one preserving almost the full length of the observation thus being affected by higher background emission but with more source counts; the second with $\sim$50\% less source counts but with higher signal-to-noise ratio. 
The variability of CID-42, already observed in the previous 15 observations, is confirmed with these new data, showing a constant flux in the full band but the hard band flux is 20\% lower than previous observations. 

The previously observed inverted {\it P-Cygni} profile is not visible in this new observation and a limit on the equivalent width ($<$162 eV at 90\% confidence level) of the absorption line is measured. The disappearance of the absorption line, detected at 3$\sigma$ in previous \xmm\ and \chandra\ observations, may be explained by the combination of several factors: high background radiation, fainter continuum flux and a change in the intrinsic properties of the inflowing material.

The analysis of extensive Monte Carlo simulations return consistent results for case A and B simulated spectra. Moreover, an EW larger than the measured upper limit should have been detected with a probability of at least 98.8\% if the line had the same intensity of the one measured in C10. 
Therefore, the higher background and faint continuum flux are not enough to explain the non-detection of the absorption line in the new spectrum, implying changes in the properties of the inflowing material. 

The simulations results indicate that the line intensity could be lower than 50\% of C10 intensity with a probability of 90\%.

CID-42 has been interpreted by C10 as a GW recoiling SMBH or as a system of 2 SMBHs moving away one from the other for slingshot effect. In this second scenario, C10 explained the X-ray absorption line as an outflow produced by the foreground active SMBH at the rearer one, which illuminates the otherwise undetectable wind. However, given the absence of X-ray emission from one of the two nuclei, C12 proposed that the most likely explanation for CID-42 is the one of a recoiling SMBH moving out from the galaxy at a velocity of $\sim$1300 km/s, thus the outflow scenario for the X-ray absorption line cannot be further supported. Detailed simulations support the recoiling scenario interpretation (Blecha et al. 2013).

In the recoiling scenario, the absorption feature is interpreted as due to absorption by ionized iron located in the proximity of the recoiling SMBH and falling with relativistic velocity into it. As the SMBH recoils from the center of the galaxy, the regions close to the SMBH, with velocities greater than the recoil velocity, are dragged along\footnote{For CID-42, all the material within 10$^5$ gravitational radii from the SMBH will move with it.} 

The equivalent width of the absorption feature in C10 implied a high column density of the
absorber (N$_H \sim$ 5 $\times$ 10$^{23}$ cm$^{-2}$), and the modest absorption in the soft band required the absorber to be
highly ionized (log $\xi \sim$3). The disappearance of both redshifted and blueshifted absorption lines is not unusual and have been seen in many examples in the literature (Tombesi et al. 2010). To justify the disappearance of the line in this work data, assuming the same ionization parameter given that no obscuration is seen in the soft band, the column density of the absorber should have now dropped by at least a factor 10 (N$_H<$ 5 $\times$ 10$^{22}$ cm$^{-2}$). Vice versa, assuming the same density of the absorber of C10, the ionization parameter should be log $\xi >$4 in order to not detect the absorption line (e.g., the gas has to be extremely ionized). 

A strong increase of ionization parameter can be generally associated to an increase of flux from the ionizing source, however, in CID-42 the X-ray flux is overall decreasing (see Fig. \ref{lightcurve}). A delay between the flux burst and the increase of ionization state is though expected depending on the distance between the cloud and the central source.  
Adopting a velocity law for the gas, as in Equation 1 of Longinotti et al. (2007), a distance of the infalling cloud from the SMBH of $\sim$100-500 gravitational radii is estimated assuming the possible range of infall velocities (0.02-0.14 $c$, depending on the iron ionization state), the BH mass and the Eddington ratio reported in C10. This distance translates into a light travel time of 4-20 $\times$10$^4$ s (in the observer frame), which is very short and implies that 
the source should have experienced a strong decrease in ionizing flux in just few tens of hours before our observation, which is highly unlikely.
A change in ionization state is then less preferred than a change in the density of the gas to explain the variability of the absorption line.

It is possible also to derive the life time of the infalling gas of 0.3--11 $\times$10$^6$ s (in the observer frame).  This lifetime is significantly lower than the time elapsed between the X-ray observations, but larger than the duration of each observation. If the infalling material is clumpy and made of discrete clouds, in each observation a different cloud is then observed, thus its properties do not need to be related with the properties of the cloud/absorption line in previous observations. If the infalling gas is instead a continuous flow, peaks with different density must occur to produce absorption lines with changing properties. 

Tombesi et al. (2010) proposed that absorption lines observed at energies below the FeK neutral line could also be interpreted as blueshifted 
transitions of helium or hydrogen like ions from elements lighter than iron (Si, S, Ar, Ca), with very high outflowing velocities (0.1-0.5$c$). 
In this scenario, given that CID-42 absorption line was observed at E$_{rest}$=5.86 keV, the outflowing velocity of the absorber would be 0.3-0.7$c$, which is very unlikely.

A redshifted absorption line with variable properties between different observations has been observed also in source E1821+643 (Yaqoob \& Serlemitsos 2005, Jim\'enez-Bail\'on et al. 2007), candidate recoiling SMBH using optical polarized spectra (Robinson et al. 2010). The redshifted absorption line in  E1821+643 has been observed in both \chandra\ and \xmm\ spectra as for CID-42.  
If a connection between the presence of variable X-ray absorption lines and a recoiling SMBH exists, it could be the subject of further investigation and modeling. 

\section{Summary}

The results on the analysis of the \xmm\ observations of the absorption line previously observed in CID-42 returns only a limit of EW$<$162 eV. 
The probability that the line intensity is not the same of C10 is 98.2\%, with a probability of 90\% that the line has now an intensity of 50\% of C10 or less. 
Therefore the non-detection of the line cannot be explained by high background or lower continuum but requires variability of the absorber properties, most likely variability of the absorber density, which is expected given the nature of these transient features.

Probing the gas kinematics and dynamics in the region closest to the SMBH is fundamental to
understand the geometry and the accretion mode of the SMBH, in particular in the interesting case of CID-42 where the SMBH is believed to move away from the center of the galaxy. New observations are indeed needed to understand the nature of this absorption line: a monitoring of the source, with multiple long ($\sim$100 ks) observations would be ideal to follow the variability of the absorbing material and of the broad-band continuum flux.

Absorption (blue and red shifted) features in X-ray data are an interesting subject for AGN studies as these are signature of high velocity material in the proximity of the SMBH, possibly responsible for mechanical feedback between SMBH and galaxy. However these features are intrinsically variable thus highly debated and difficult to observe at the limit of the capability of CCD and gratings (only for few local sources) instruments. Future missions as Athena+ could open a new window on this subject allowing to perform more detailed modeling and collect statistical samples of these features (Georgakakis et al. 2013).

\acknowledgments
The authors thanks A. Longinotti for useful discussions and the anonymous referee for the useful suggestions to improve the overall 
interpretation of this source.
F.C., S.M. and G.L. acknowledge support by the NASA 
contract 11-ADAP11-0218. K.J. acknowledges support by the German Science Foundation (DFG), grant Ja 1114/3-1. 
S.K. research was supported by the DFG cluster of excellence
''Origin and Structure of the Universe`` (www.universe$-$cluster.de).
M.E. and S.K. thank the Aspen Center for Physics for support and hospitality. The
Aspen Center for Physics is supported by the National Science Foundation under
Grant No. PHYS-1066293. A.C. acknowledges financial contribution from the agreement ASI-INAF I/009/10/0 and INAF-PRIN 2011


\begin{thebibliography}{}
\bibitem[Arnaud(1999)]{1999HEAD....4.3301A} Arnaud, K.~A.\ 1999, Bulletin 
of the American Astronomical Society, 31, 734 


\bibitem[Batcheldor et al.(2010)]{2010ApJ...717L...6B} Batcheldor, D., 
Robinson, A., Axon, D.~J., Perlman, E.~S., 
\& Merritt, D.\ 2010, \apjl, 717, L6 


\bibitem[Bekenstein(1973)]{1973ApJ...183..657B} Bekenstein, J.~D.\ 1973, 
\apj, 183, 657 


\bibitem[Blecha et al.(2013)]{2013MNRAS.428.1341B} Blecha, L., Civano, F., 
Elvis, M., \& Loeb, A.\ 2013, \mnras, 428, 1341 


\bibitem[Blecha et al.(2011)]{2011MNRAS.412.2154B} Blecha, L., Cox, T.~J., 
Loeb, A., \& Hernquist, L.\ 2011, \mnras, 412, 2154 


\bibitem[Bonning et al.(2007)]{2007ApJ...666L..13B} Bonning, E.~W., 
Shields, G.~A., \& Salviander, S.\ 2007, \apjl, 666, L13 


\bibitem[Brusa et al.(2010)]{2010ApJ...716..348B} Brusa, M., Civano, F., 
Comastri, A., et al.\ 2010, \apj, 716, 348 


\bibitem[Cappelluti et 
al.(2009)]{2009A&A...497..635C} Cappelluti, N., Brusa, M., Hasinger, G., et al.\ 2009, \aap, 497, 635 


\bibitem[Cappelluti et al.(2007)]{2007ApJS..172..341C} Cappelluti, N., 
Hasinger, G., Brusa, M., et al.\ 2007, \apjs, 172, 341 


\bibitem[Cappi et 
al.(2009)]{2009A&A...504..401C} Cappi, M., Tombesi, F., Bianchi, S., et al.\ 2009, \aap, 504, 401 


\bibitem[Cash(1979)]{1979ApJ...228..939C} Cash, W.\ 1979, \apj, 228, 939 


\bibitem[Civano et al.(2012)]{2012ApJS..201...30C} Civano, F., Elvis, M., 
Brusa, M., et al.\ 2012, \apjs, 201, 30 


\bibitem[Civano et al.(2012)]{2012ApJ...752...49C} Civano, F., Elvis, M., 
Lanzuisi, G., et al.\ 2012, \apj, 752, 49 


\bibitem[Civano et al.(2010)]{2010ApJ...717..209C} Civano, F., Elvis, M., 
Lanzuisi, G., et al.\ 2010, \apj, 717, 209 


\bibitem[Colpi 
\& Dotti(2009)]{2009arXiv0906.4339C} Colpi, M., \& Dotti, M.\ 2009, arXiv:0906.4339 


\bibitem[Dadina et 
al.(2005)]{2005A&A...442..461D} Dadina, M., Cappi, M., Malaguti, G., Ponti, G., \& de Rosa, A.\ 2005, \aap, 442, 461 


\bibitem[Elvis(2000)]{2000ApJ...545...63E} Elvis, M.\ 2000, \apj, 545, 63


\bibitem[Elvis et al.(2009)]{2009ApJS..184..158E} Elvis, M., Civano, F., 
Vignali, C., et al.\ 2009, \apjs, 184, 158 


\bibitem[Eracleous et al.(2012)]{2012ApJS..201...23E} Eracleous, M., 
Boroson, T.~A., Halpern, J.~P., \& Liu, J.\ 2012, \apjs, 201, 23 


\bibitem[Fiore et 
al.(2012)]{2012A&A...537A..16F} Fiore, F., Puccetti, S., Grazian, A., et al.\ 2012, \aap, 537, A16 


\bibitem[Georgakakis et al.(2013)]{2013arXiv1306.2328G} Georgakakis, A., 
Carrera, F., Lanzuisi, G., et al.\ 2013, arXiv:1306.2328 


\bibitem[Grupe et al.(2012)]{2012ApJS..199...28G} Grupe, D., Komossa, S., 
Gallo, L.~C., et al.\ 2012, \apjs, 199, 28 


\bibitem[Guedes et al.(2011)]{2011ApJ...729..125G} Guedes, J., Madau, P., 
Mayer, L., \& Callegari, S.\ 2011, \apj, 729, 125 


\bibitem[Hopkins et al.(2008)]{2008ApJS..175..356H} Hopkins, P.~F., 
Hernquist, L., Cox, T.~J., \& Kere{\v s}, D.\ 2008, \apjs, 175, 356 


\bibitem[Jim{\'e}nez-Bail{\'o}n et 
al.(2007)]{2007A&A...461..917J} Jim{\'e}nez-Bail{\'o}n, E., Santos-Lle{\'o}, M., Piconcelli, E., et al.\ 2007, \aap, 461, 917 


\bibitem[Jonker et al.(2010)]{2010MNRAS.407..645J} Jonker, P.~G., Torres, 
M.~A.~P., Fabian, A.~C., et al.\ 2010, \mnras, 407, 645 


\bibitem[Kalberla et 
al.(2005)]{2005A&A...440..775K} Kalberla, P.~M.~W., Burton, W.~B., Hartmann, D., et al.\ 2005, \aap, 440, 775


\bibitem[King 
\& Pounds(2003)]{2003MNRAS.345..657K} King, A.~R., \& Pounds, K.~A.\ 2003, \mnras, 345, 657 


\bibitem[Komossa(2012)]{2012AdAst2012E..14K} Komossa, S.\ 2012, Advances in 
Astronomy, 2012,  


\bibitem[Komossa et al.(2008)]{2008ApJ...678L..81K} Komossa, S., Zhou, H., 
\& Lu, H.\ 2008, \apjl, 678, L81 


\bibitem[Lanzuisi et al.(2013)]{2013MNRAS.431..978L} Lanzuisi, G., Civano, 
F., Elvis, M., et al.\ 2013, \mnras, 431, 978 


\bibitem[Loeb(2007)]{2007PhRvL..99d1103L} Loeb, A.\ 2007, Physical Review 
Letters, 99, 041103 


\bibitem[Longinotti et al.(2007)]{2007MNRAS.374..237L} Longinotti, A.~L., 
Sim, S.~A., Nandra, K., \& Cappi, M.\ 2007, \mnras, 374, 237 


\bibitem[Matt et 
al.(2005)]{2005A&A...435..857M} Matt, G., Porquet, D., Bianchi, S., et al.\ 2005, \aap, 435, 857 


\bibitem[Merritt et al.(2006)]{2006MNRAS.367.1746M} Merritt, D., 
Storchi-Bergmann, T., Robinson, A., et al.\ 2006, \mnras, 367, 1746


\bibitem[Nandra et al.(1999)]{1999ApJ...523L..17N} Nandra, K., George, 
I.~M., Mushotzky, R.~F., Turner, T.~J., 
\& Yaqoob, T.\ 1999, \apjl, 523, L17 


\bibitem[Nandra et al.(2007)]{2007MNRAS.382..194N} Nandra, K., O'Neill, 
P.~M., George, I.~M., \& Reeves, J.~N.\ 2007, \mnras, 382, 194 


\bibitem[O'Neill et al.(2007)]{2007MNRAS.381L..94O} O'Neill, P.~M., Nandra, 
K., Cappi, M., Longinotti, A.~L., \& Sim, S.~A.\ 2007, \mnras, 381, L94 


\bibitem[Peres(1962)]{1962PhRv..128.2471P} Peres, A.\ 1962, Physical 
Review, 128, 2471 


\bibitem[Ponti et 
al.(2013)]{2013A&A...549A..72P} Ponti, G., Cappi, M., Costantini, E., et al.\ 2013, \aap, 549, A72 


\bibitem[Pounds 
\& Reeves(2009)]{2009MNRAS.397..249P} Pounds, K.~A., \& Reeves, J.~N.\ 2009, \mnras, 397, 249 


\bibitem[Reeves et al.(2005)]{2005ApJ...633L..81R} Reeves, J.~N., Pounds, 
K., Uttley, P., et al.\ 2005, \apjl, 633, L81 


\bibitem[Robinson et al.(2010)]{2010ApJ...717L.122R} Robinson, A., Young, 
S., Axon, D.~J., Kharb, P., \& Smith, J.~E.\ 2010, \apjl, 717, L122 


\bibitem[Shields et al.(2009)]{2009ApJ...696.1367S} Shields, G.~A., 
Bonning, E.~W., \& Salviander, S.\ 2009, \apj, 696, 1367 


\bibitem[Sijacki et al.(2011)]{2011MNRAS.414.3656S} Sijacki, D., Springel, 
V., \& Haehnelt, M.~G.\ 2011, \mnras, 414, 3656 


\bibitem[Steinhardt et al.(2012)]{2012ApJ...759...24S} Steinhardt, C.~L., 
Schramm, M., Silverman, J.~D., et al.\ 2012, \apj, 759, 24 


\bibitem[Tombesi et al.(2013)]{2013MNRAS.430.1102T} Tombesi, F., Cappi, M., 
Reeves, J.~N., et al.\ 2013, \mnras, 430, 1102 


\bibitem[Tombesi et al.(2010)]{2010A&A...521A..57T} Tombesi, F., Cappi, M., Reeves, J.~N., et al.\ 2010, \aap, 521, A57 


\bibitem[Vaughan 
\& Uttley(2008)]{2008MNRAS.390..421V} Vaughan, S., \& Uttley, P.\ 2008, \mnras, 390, 421 


\bibitem[Volonteri et al.(2003)]{2003ApJ...582..559V} Volonteri, M., 
Haardt, F., \& Madau, P.\ 2003, \apj, 582, 559 


\bibitem[Yaqoob 
\& Serlemitsos(2005)]{2005ApJ...623..112Y} Yaqoob, T., \& Serlemitsos, P.\ 2005, \apj, 623, 112 


\end{thebibliography}
\end{document}